\documentstyle[twocolumn,aps,prl,array]{revtex}
\def\s#1{\slash\!\!\!{#1}}

\def\bee{\begin{eqnarray}}
\def\eee{\end{eqnarray}}
\def\nn{\nonumber\\}

\def\fr#1#2{\frac{#1}{#2}}

\begin{document}
\draft
\title{On the NLO Power Correction to Photon-Pion Transition Form Factor}
\author{Tsung-Wen Yeh\thanks{E-mail: twyeh@cc.nctu.edu.tw}}
\address{Institute of Physics, National Chiao-Tung University,
Hsinchu, Taiwan 300, Republic of China}
\maketitle
\begin{abstract}
We propose a perturbative evaluation for the next-to-leading-order (NLO) $O(1/Q^4)$ power correction to the photon pion transition form factor $F_{\pi \gamma}(Q^2)$. The effects of the NLO power correction are analyzed.   
\end{abstract}
\narrowtext
\section{INTRODUCTION}
The photon-pion transition process $\gamma^*\pi\to\gamma$ provides a good example for tests of QCD. This is because, at tree level in PQCD, there only involve electro-magnetic interactions. The short distance interactions can be calculated in a definite way.  The amplitude for the process $\gamma^*\pi\to\gamma$ can be expressed as $M(\gamma^*\pi\to\gamma)=-ie^2\epsilon_{\mu\alpha\beta\lambda}P_1^{\alpha}P_2^{\beta}\epsilon^{\lambda}F_{\pi\gamma}(Q^2)$, where $P_1$ means the pion momentum, $P_2$ and $\epsilon^{\lambda}$ denote the momentum and polarization of the real photon and $Q^2=-(P_2-P_1)^2$ is the virtuality of the virtual photon. The QCD dynamics are contained in the form factor $F_{\pi\gamma}(Q^2)$. In PQCD, the leading form factor $F_{\pi\gamma}(Q^2)$ can be expressed as 
\bee
F_{\pi\gamma}(Q^2)=4C_\pi\int_0^1 dx \frac{\phi_2(x)}{Q^2x(1-x)}
\eee
with $C_\pi$ the charge  factor and $\phi_2(x)$ the pion wave function.
In the high energy limit $Q^2\to\infty$, the asymptotic of the form factor $F_{\pi\gamma}$ can be evaluated and have an extremely simple form \cite{BL1}
\begin{eqnarray}
F_{\pi\gamma}(Q^2)\Bigg|_{Q^2\to\infty}=\frac{2f_\pi}{Q^2}
\end{eqnarray}
that it is only in terms of the pion decay constant, $f_\pi=93$ MeV and $Q^2$. However, the above asymptotic is about $15\%$ lower than the upper end of the experimental data  \cite{CLEO}. There have been many explanations about this difference between the theory and the experiment. For expample, the inclusion of NLO correction in $\alpha_s$ \cite{Kroll:1996jx} and the transverse structure of the pion wave function \cite{pp1,pp2}.  It has also been shown \cite{CLEO} that the data can also be described by the Brodsky-Lepage interpolating formula \cite{BL2} for both $Q^2\to\infty$ and $Q^2\to 0$ of $F_{\pi\gamma}(Q^2)$
\begin{eqnarray}
F_{\pi\gamma}^{BL}(Q^2)=\frac{2f_\pi}{s_0+Q^2}
\end{eqnarray}
where $s_0=8\pi^2 f_\pi^2\approx 0.68$ GeV$^2$. This implies that the NLO power correction, the $O(1/Q^4)$ correction, might be important. 

In this paper, we shall present a perturbative approach to calculate the NLO power correction for form factor $F_{\pi \gamma}(Q^2)$. The calculations are related to the terminology of the collinear expansion \cite{CL,twy1}. The expansion has the following features:(1) it preserves individual gauge invariance of the soft function and the hard function;(2) it can systematically separate the leading twist (LT) contributions from the next-to-leading twist (NLT) contributions ;(3) it encounters the high twist contributions from non-collinear partons, wrong spin projection and higher Fock states;(4) it is a twist-by-twist expansion and free from twist mixing problem;(5) it is a Feynman diagram approach such that the parton picture for high twist contributions is preserved. 

Our main results are summarized as follows. We shall employ collinear expansion to derive the LT and NLT contributions for tree diagrams of processes $\gamma^*\pi\to\gamma$. The NLT contributions to $F_{\pi\gamma}(Q^2)$ involve four NLT pion DAs. Two of them come from non-vanishing masses of the valence quarks. By the help of equations of motion, the four DAs are reduced to two DAs. The asymptotic forms for the remaining DAs are employed. The theory appears in a good agreement with the data.    

The organization of the remaining text is as follows. We sketch the collinear expansion for the process $\gamma^*\pi\to\gamma$ in Sec.~II. The explicit formula up to order $O(Q^{-4})$ of the process $\gamma^*\pi\to\gamma$ are given in Sec.~III. Sec.~IV is devoted to conclusions.

\section{Collinear Expansion and Factorization}

We sketch the procedures of performing collinear expansion for $\gamma^*\pi\to\gamma$.

\subsection{Tree Level Collinear Expansion}
Let $\sigma=\sigma_p(k)\otimes\phi(k)$ represent the lowest order amplitude for $\gamma^*(q)\pi(P_1)\to\gamma(P_2)$ as depicted in Fig.~1(a) and (b).  The $\sigma_p(k)$ denotes the amplitude for partonic subprocess and the $\phi(k)$ represents the meson DA. The $\otimes$ means convolution integral over the loop momentum $k$ and the traces over the color indices and spin indices. If we assign the momentum of the final state photon in the minus light-cone direction, $P_2=Q^2 n/2$, and the momentum of the light meson in the plus light-cone direction, $P_1=p$, then the leading configuration for the process can be a function of the collinear momentum $k=x p$ with $x=k\cdot n$ the fraction of the momentum of the meson carried by the partons. $p$ and $n$ represent light-like vectors in $+$ and $-$ directions and satisfy $n\cdot p=1$. The first step is to perform a Taylor expansion for the parton amplitude 
\bee\label{st1}
\sigma_p(k)=\overline{\sigma}_p(k=x p)
                  +{(\overline{\sigma}_p)_{\alpha}(x,x) }w^\alpha_{\alpha^\prime}k^{\alpha^\prime}+\cdots
\eee
where we have assumed the low energy theorem 
\bee
\fr{\partial}{\partial k^\alpha}\sigma_p(k)\Bigg|_{k=x p}=(\overline{\sigma}_p)_{\alpha}(x,x)\ . 
\eee  
and have employed $w^\alpha_{\alpha^\prime}k^{\alpha^\prime}=(k-xp)^\alpha$ and $w^\alpha_{\alpha^\prime}=g^\alpha_{\alpha^\prime}-p^\alpha n_{\alpha^\prime}$.
The leading term $\overline{\sigma}_p\otimes \phi$ contains leading, next-to-leading and higher twist contributions in accord with the spin structures of the leading parton amplitude $\overline{\sigma}_p$. That is
those terms are proportional to $\s{n}$ or $\s{p}$. The terms proportional to $\s{n}$ would project collinear $q\bar{q}$ pair from the meson, while those terms proportional to $\s{p}$ would not diminish only when the $q\bar{q}$ pair carry non-collinear momenta. The second step is to substitute the leading parton amplitude $\overline{\sigma}_p$ into the convolution integral with the meson wave function $\phi$ to extract the NLT contribution $\overline{\sigma}_p\otimes\phi_1$ from the leading one $\overline{\sigma}_p\otimes\phi_0$
    \bee\label{st2}
     \overline{\sigma}_p\otimes \phi=\overline{\sigma}_p\otimes\phi_0 +
     \overline{\sigma}_p\otimes\phi_1 +\cdots ,
    \eee
where $\phi_0$ and $\phi_1$ denote the leading and next-to-leading meson DAs, respectively. The high twist DA $\phi_1$ contains both short distance and long distance contributions. The short distance contributions of $\phi_1$ arise from the non-collinear components of $k$. By the equations of motion, the non-collinear components of $k$ will induce one quark-gluon vertex $i\gamma_\alpha$ and one special propagator $i\s{n}/2 x_i$ \cite{CL}. Because that the special propagator is not propagating, the quark-gluon vertex and the special propagator should be included into the hard function, $\overline{\sigma}_p$. In this way, we may factorize $\phi_1$ as $\phi_1\approx(\phi_1^H)_\alpha w^\alpha_{\alpha^\prime}(\phi_1^S)^{\alpha^\prime}$ and absorb the short distance piece $(\phi_1^H)_\alpha$ into $\overline{\sigma}_p$. It leads to the third step 
\bee\label{st3}
     \overline{\sigma}_p\otimes\phi_1&=&\overline{\sigma}_p
        \otimes((\phi_1^H)_\alpha\bullet w^\alpha_{\alpha^\prime} (\phi_1^S)^{\alpha^\prime})\nn
      &=&(\overline{\sigma}_p\bullet\phi_1^H)_\alpha\otimes w^\alpha_{\alpha^\prime} (\phi_1^S)^{\alpha^\prime}\ ,
\eee
where  $(\phi_1^S)^{\alpha^\prime}$ containing covariant derivative $D^{\alpha^\prime}=i\partial^{\alpha^\prime}-g A^{\alpha^\prime}$ is implied.
Notice that the light-cone gauge $n\cdot A=0$ assures  $w^\alpha_{\alpha^\prime}A^{\alpha^\prime}=A^\alpha$.
The second term of Eq.(\ref{st1}) also contribute to high twist corrections, as it convolutes with $\phi_0$. The momentum factor $k^\alpha$ will be absorbed by $\phi_0$ to become a coordinate derivative, denoted as  $k^\alpha\phi_0\equiv \phi_{1,\partial}^\alpha$. Consider another NLT contributions $\sigma_1\approx(\overline{\sigma}_p)_\alpha\otimes w^\alpha_{\alpha^\prime}\phi_{1,A}^{\alpha^\prime}$ from Fig.~1(c) and (d), where $\phi_{1,A}^{\alpha}$ contains gauge fields. We have employed the approximation that $(\overline{\sigma}_p)_\alpha\otimes \phi_{1,A}^{\alpha}$ is the leading term of $\sigma_1$. This comes to the fourth step: 
\bee\label{st4}
     (\overline{\sigma}_p)_{\alpha}\otimes w^\alpha_{\alpha^\prime} \phi_{1,\partial}^{\alpha^\prime}
+ (\overline{\sigma}_p)_\alpha\otimes w^\alpha_{\alpha^\prime}\phi_{1,A}^{\alpha\prime}\equiv (\overline{\sigma}_p)_1\otimes (\phi_1^S)\ ,
\eee
where we have employed $\phi_{1,\partial}^{\alpha}+\phi_{1,A}^\alpha=(\phi_1^S)^\alpha$. However, it will be found that  $(\overline{\sigma}_p)_1$ diminishes as convoluting with twist-4 DA $\phi_1^\Gamma$ (see below definition).  
Up to NLO, we may drop the $(\overline{\sigma}_p)_1$ term and arrive at the factorization for tree amplitudes
\bee
\sigma_0 + \sigma_1\approx \overline{\sigma}_p\otimes\phi_0 +(\overline{\sigma}_p\bullet\phi_1^H)\otimes \phi_1
\eee
where $\phi_1$ means $\phi_1^S$.
There involves only one NLT DA $\phi_1$ for NLO power corrections.

To proceed, we need to consider the factorizations of the spin indices, the color indices and the momentum integrals over loop partons. For factorization of spin indices, we employ the expansion of the meson DA into its spin components as
\bee
\phi_{0,1} &=&\sum_{\Gamma}\phi_{0,1}^\Gamma \Gamma
\eee
where $\Gamma$ means Dirac matrix $\Gamma=1,\gamma^\mu,\gamma^\mu\gamma_5,\sigma^{\mu\nu}$. 
The factorization of the color indices take the convention that the color indices of the parton amplitudes are extract and attributed to the meson DAs. The factorization of the momentum integral is performed by making use of the fact that the leading parton amplitudes depend only on the momentum fraction variables $x_i$. The identity can always be used
\bee
\int^1_0 dx_i\delta(x_i-k_i\cdot n)=1\ .
\eee

The choice of the lowest twist components $\phi_{0,1}^\Gamma$ of $\phi_{0,1}$ is made by employing the power counting. 
Consider $\pi$ meson whose high twist DA $\phi^{\mu_1\cdots\mu_F;\alpha_1\cdots\alpha_B}$ has the fermion index $F$ and the boson index $B$. The fermion index $F$ arise from the spin index factorization for $2F$ fermion lines connecting DA and parton amplitude and the boson index $B$ denotes the $n_D$ power of  momenta in previous collinear expansion and the $n_G$ gluon lines as $B=n_D+n_G$. We may write
\bee\label{pc1}
\phi^{\mu_1\cdots\mu_F;\alpha_1\cdots\alpha_B}=\sum_i \Lambda^{\tau_i-1}e^{\mu_1\cdots\mu_F;\alpha_1\cdots\alpha_B}_i\phi^i
\eee
where $\Lambda$ denotes a small scale associated with DA. Spin polarizers $e_i$ denote the combination of  vectors $p^\mu$, $n^\mu$ and $\gamma^\mu_\perp$. Variable $\tau_i$ represents the twist of DA $\phi^i$. The restrictions over projector $e^{\mu_1\cdots\mu_F;\alpha_1\cdots\alpha_B}_i$ are 
\bee\label{pc2}
n_{\alpha_j}e^{\mu_1\cdots\mu_F;\alpha_1\cdots\alpha_j\cdots\alpha_B}_i=0
\eee
which are due to the fact that polarizers $e_i$ are always projected by  $w^{\alpha}_{\alpha^\prime}$.
The dimension of $\phi^{\mu_1\cdots\mu_F;\alpha_1\cdots\alpha_B}$ is determined by dimensional analysis
\bee\label{pc3}
d(\phi)=3F+B-1
\eee 
By equating the dimensions of both sides of Eq.(\ref{pc1}), one can derive the minimum of $\tau_i$ 
\bee\label{pc4}
\tau_i^{\min} =2F+B+\frac{1}{2}[1-(-1)^B]\ .
\eee
It is obvious from Eq.(\ref{pc4}) that there are only finite numbers of fermion lines, gluon lines and derivatives contributes to a given power of $1/Q^2$.

\subsection{Collinear Expansion for Arbitrary Loop Orders}
The factorization theorem for the NLO power correction should be proven in order to have a confident PQCD formalism. Before this can be done, we can at least show that the collinear expansion is compatible with the conventional approach for proving the factorization theorem. The conventional approach is based on the factorization of the soft divergences from the ultraviolate divergences arising from the radiative corrections. The soft divergences are shown to be cancelled or absorbed by the meson wave function such that the parton amplitude is free from the soft divergences. The ultraviolate divergences can be absorbed by the parton amplitude. We shall assume that the factorization of the soft and ultrviolate divergences can be done upto NLO power correction. Then, we can show that the the collinear expansion for tree diagrams can be straightforwardly extended to those diagrams containing arbitrary loop corrections. The starting point is to notice that the collinear expansion for the one loop corrections in the collinear region of the radiative gluons can be written down as
\bee\label{cla1}
&&(\sigma_0^{(0)} + \sigma_1^{(0)}+\sigma_0^{(1)} + \sigma_1^{(1)})\Bigg|_{\makebox{\tiny collinear gluons}}\nn
&\approx& \sum_{j=0,1} \overline{\sigma}_p^{(0)}\otimes\phi_0^{(j)}
+\overline{\sigma}_p^{(0)}\bullet(\phi_1^H)^{(0)}\otimes \phi_1^{(j)}\nn
\eee
where superscript $(0),(1)$ denote tree and one loop corrections, respectively. This is because, as the collinear gluons with momentum $l\sim (Q,\lambda^2/Q,\lambda)$ go through the fermion lines, the valence fermion momenta behave similarly to those in the tree level expansion. This leads to the fact that the collinear expansions for one loop amplitudes in collinear region can be performed just like for tree amplitudes. The soft gluon corrections can not affect the collinear expansion. The cancellations of double logarithms are assured in light-cone gauge by adding ladder and self energy diagrams. The one loop corrected parton amplitudes are determined by subtracting the amplitudes in collinear and soft regions from the full one loop amplitudes . Following the standard considerations \cite{twy2}, the LT parton amplitude $\overline{\sigma}_p^{(0)}$ and NLT parton amplitude $\overline{\sigma}_p^{(0)}\bullet(\phi_1^H)^{(0)}$ are infrared finite, and the soft divergences are absorbed by $\phi_0^{(1)}$ and $\phi_1^{(1)}$. The one loop factorization is derived up to NLT order
\bee\label{cla2}
&&(\sigma_0^{(0)} + \sigma_1^{(0)}+\sigma_0^{(1)} + \sigma_1^{(1)})\nn
&\approx& [\sum_{i=0}^1 \overline{\sigma}_p^{(i)}]\otimes[\sum_{j=0}^1\phi_0^{(j)}]\nn
&&+[\sum_{i=0}^1 \sum_{j=0}^i\overline{\sigma}_p^{(j)}\bullet(\phi_1^H)^{(i-j)}]\otimes [\sum_{k=0,1} \phi_1^{(k)}]\ .
\eee
The generalization to arbitrary loop orders can be obtained by iteration.
Suppose that 
\bee
\sigma=(\sigma_p)_0\otimes\phi_0+(\sigma_p)_1\otimes\phi_1
\eee
where
\bee
\sigma&=&\sum_{i=0}^N \sigma^{(i)}\ , 
(\sigma_p)_0=\sum_{i=0}^N (\overline{\sigma}_p)_0^{(i)}\ ,
\phi_{0,1}=\sum_{i=0}^N \phi_{0,1}^{(i)}\nn
(\sigma_p)_1&=&\sum_{i=0}^N (\sigma_p)_1^{(i)}\ ,
\eee
where
\bee
(\sigma_p)_1^{(i)}=\sum_{j=0}^N \overline{\sigma}_p^{(j)}\bullet(\phi_1^H)^{(i-j)}\ .
\eee
The above factorization still holds for $N+1$ order corrections, since the collinear gluons cannot attach to the parton amplitudes. The remaining proof of factorization requires the cancellations of double logarithms of soft divergences, the single soft logarithms absorbed by pion DAs and the infrared finiteness of the parton amplitudes. This can be achieved by standard analysis (see e.g. \cite{BL1}) and it is left to other publish \cite{twy2}. 

\section{$O(1/Q^4)$ CONTRIBUTIONS OF $\gamma^*\pi\to\gamma$}
The lowest-order diagrams are displayed in Fig.~1. 
By applying the previous collinear expansion to separate the LT and NLT contributions in a factorized form, we may write the result as
\begin{eqnarray}\label{pi3}
M(\gamma^*\pi\to\gamma)=-ie^2\epsilon_{\mu\alpha\beta\lambda}P_1^{\alpha}P_2^{\beta}\epsilon^{\lambda}F_{\pi\gamma}(Q^2)\ ,
\end{eqnarray}
where $\epsilon^{\lambda}$ denotes the polarization vector of the final state photon. The leading order of $F_{\pi\gamma}(Q^2)$ is calculated from Fig.1(a) and (b) as
\begin{eqnarray}
F_{\pi\gamma}^{\makebox{\tiny LO}}(Q^2)=4C_\pi\int_0^1 dx \frac{\phi_2(x)}{Q^2x(1-x)}\ ,
\end{eqnarray}
where the charge factor $C_\pi=(e_u^2-e_d^2)/\sqrt{2}$. $e_u$ and $e_d$ mean the charges of $u$ and $d$ quark in unites of the elementary charge.
The NLO of $F_{\pi\gamma}(Q^2)$ is evaluated from Fig.2
\begin{eqnarray}
F_{\pi\gamma}^{\makebox{\tiny NLO}}(Q^2)=-16C_\pi\int_0^1 dx \frac{[G(x)+\tilde{G}(x)(1-2x)]}{Q^4x(1-x)}\ .
\end{eqnarray}
The relevant DAs are expressed explicitly as follows
\begin{eqnarray}
\phi_2(x)&=&-i\frac{1}{4}\int_0^\infty \frac{d\lambda}{(2\pi)}e^{i\lambda x}
\langle 0|\bar{q}(0)\gamma_5 \not{n}q(\lambda n)|\pi(P_1)\rangle\\
G(x)&=&-\frac{1}{8}\epsilon_\perp^{\alpha\beta}\int_0^1 d x_1\int_0^\infty \frac{d\lambda}{(2\pi)}\frac{d\eta}{(2\pi)}e^{i\eta( x_1-x)}
e^{i\lambda x}\nn
&&\times\langle 0|\bar{q}(0)\gamma_{\alpha}D_{\beta}(\eta n)q(\lambda n)|\pi(P_1)\rangle\ ,\\
\tilde{G}(x)&=&-\frac{i}{8}d_\perp^{\alpha\beta}\int_0^1 d x_1\int_0^\infty \frac{d\lambda}{(2\pi)}\frac{d\eta}{(2\pi)}e^{i\eta( x_1-x)}
e^{i\lambda x}\nn
&&\times\langle 0|\bar{q}(0)\gamma_5\gamma_{\alpha}D_{\beta}(\eta n)q(\lambda n)|\pi(P_1)\rangle\ .
\end{eqnarray}
The tensors $\epsilon_\perp^{\alpha\beta}$ and $d_\perp^{\alpha\beta}$ are defined as $\epsilon_\perp^{\alpha\beta}=\epsilon^{\alpha\beta\gamma\lambda}p_{\gamma}n_{\lambda}$ and $d_\perp^{\alpha\beta}=p^{\alpha}n^{\beta}+n^{\alpha}p^{\beta}-g^{\alpha\beta}$.

The nonvanishing valence quark mass can also contribute to NLO correction. We use the scheme that the partons involved in the hard function are massless. This does not affect the final conclusion. By assigning new contributions from the quark mass operator $m$, we get the result 
\begin{eqnarray}
&&\left.F_{\pi\gamma}^{\makebox{\tiny NLO}}(Q^2)\right|_{m_q\not= 0}\nonumber\\
&=&-8C_\pi\int_0^1 dx \frac{[H(x) +\tilde{H}(x)(1-2x)]}{Q^4x(1-x)}
\end{eqnarray}
where two twist-4 DAs are introduced
\begin{eqnarray}
H(x)&=&-\frac{i}{16}\epsilon^{\alpha\beta}_{\perp}\int_0^\infty \frac{d\lambda}{(2\pi)}
e^{i\lambda x}
\langle 0|\bar{q}(0){\it m}\sigma_{\alpha\beta}q(\lambda n)|\pi(P_1)\rangle\ ,\nn\\
\tilde{H}(x)&=&-\frac{i}{4}\int_0^\infty \frac{d\lambda}{(2\pi)}
e^{i\lambda x}
\langle 0|\bar{q}(0){\it m}\gamma_5q(\lambda n)|\pi(P_1)\rangle\ .\nn
\end{eqnarray}
Note that $H(x)$ and $\tilde{H}(x)$ are related to the conventional twist-3 pion DA $\phi_\sigma(x)$ and $\phi_p(x)$ \cite{Braun} by a factor $m_0$, the average quark mass.

The DAs $G,\tilde{G},H$ and $\tilde{H}$ are dependent and reducible under equations of motion to 
\bee
G^\prime(x)&=&-\frac{1}{16}\epsilon_\perp^{\alpha\beta}\int_0^1 d x_1\int_0^\infty \frac{d\lambda}{(2\pi)}\frac{d\eta}{(2\pi)}e^{i\eta( x_1-x)}
e^{i\lambda x}\nn
&&\times\langle 0|\bar{q}(0)\gamma_{\alpha}\s{D}(\eta n)\gamma_{\beta}q(\lambda n)|\pi(P_1)\rangle\ ,
\eee
and 
\bee
\tilde{G}^\prime(x)&=&-\frac{i}{16}d_\perp^{\alpha\beta}\int_0^1 d x_1\int_0^\infty \frac{d\lambda}{(2\pi)}\frac{d\eta}{(2\pi)}e^{i\eta( x_1-x)}
e^{i\lambda x}\nn
&&\times\langle 0|\bar{q}(0)\gamma_5\gamma_{\alpha}\s{D}(\eta n)\gamma_{\beta}q(\lambda n)|\pi(P_1)\rangle\ .
\eee
Due to the factor $1-2x$ for $\tilde{G}^\prime$, $G^\prime$ becomes dominate. The normalization of $\phi_2(x)$ is fixed from process $\pi\to\mu\nu$ such that $\phi_2^{\makebox{\tiny AS}}(x)=3f_\pi x(1-x)/\sqrt{2}$ for asymptotic (AS) model and $\phi_2^{\makebox{\tiny CZ}}(x)=15 f_\pi x(1-x)(1-2x)^2/\sqrt{2}$ for Chernyak-Zhitnitsky (CZ) model \cite{CZ}. Similarly, the normalization of $G^\prime(x)$ is determined from the axial anomaly $\pi\to 2\gamma$ \cite{BL2} to yield $G^{\prime\makebox{\tiny AS}}(x)=3\sqrt{2}\pi^2 f_\pi^3 x(1-x)$ for AS model and $G^{\prime\makebox{\tiny CZ}}(x)=15\sqrt{2}\pi^2 f_\pi^3 x(1-x)(1-2x)^2$ for CZ model. The comparison between the experiment and our result indicates that the data is in more favor of AS model than of CZ model (see Fig.3).  This is consistent with that conclusion made by Jakob {\it et al.} \cite{pp1}.
\section{Conclusions}
We have shown that the collinear expansion for $\gamma^*\pi\to\gamma$ can be systematically performed.
The $O(Q^{-4})$ power correction for $F_{\pi\gamma}(Q^2)$ has been evaluated in terms of four twist-4 DAs. The effects of NLO power correction are estimated to account for the data \cite{CLEO}. Applications of the collinear expansion to other processes can be straightforwardly perfromed.

The other sources of power correction may also be important, such as the renormalon. The investigation of this kind of power correction is beyond the scope of this paper.

We have also limited ourselves to the tree level. The factorization theorem for the NLO power correction should be proven if we require a confident PQCD formalism. As we have shown, the collinear expansion is compatible with the conventional approach for proving the factorization theorem. The collinear expansion can be performed order by order for radiative corrections.

\noindent
{\bf Acknowledgments:}
This work was supported in part by the National
Science Council of R.O.C. under the Grant No. NSC89-2811-M-009-0024.

Figure Caption
\begin{list}{}
\item Fig.1 The leading order diagrams for $\gamma^*\pi\to\gamma$. The cross symbol means the vertex of the virtual photon.
\item Fig.2 The next-to-leading-twist (NLT) diagrams for $\gamma^*\pi\to\gamma$. The propagator with one bar means the special propagator.
\item Fig.3 The pion-photon transition form factor $Q^2 F_{\pi\gamma}(Q^2)$ calculated with different distribution amplitudes: the asymptotic (solid line) and Cheryak-Zhitnitsky (dash line). The experimental data are taken from \cite{CLEO}.
\end{list}
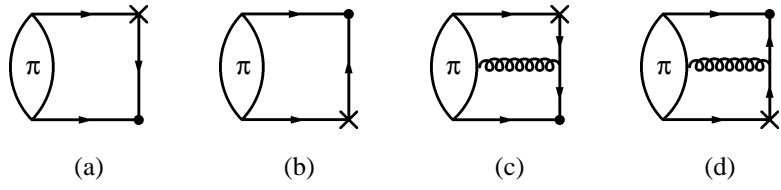
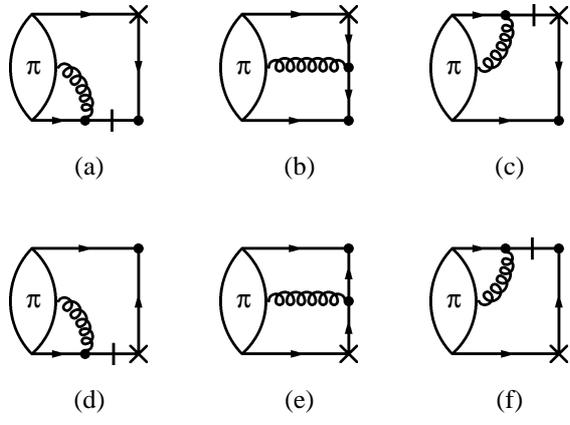
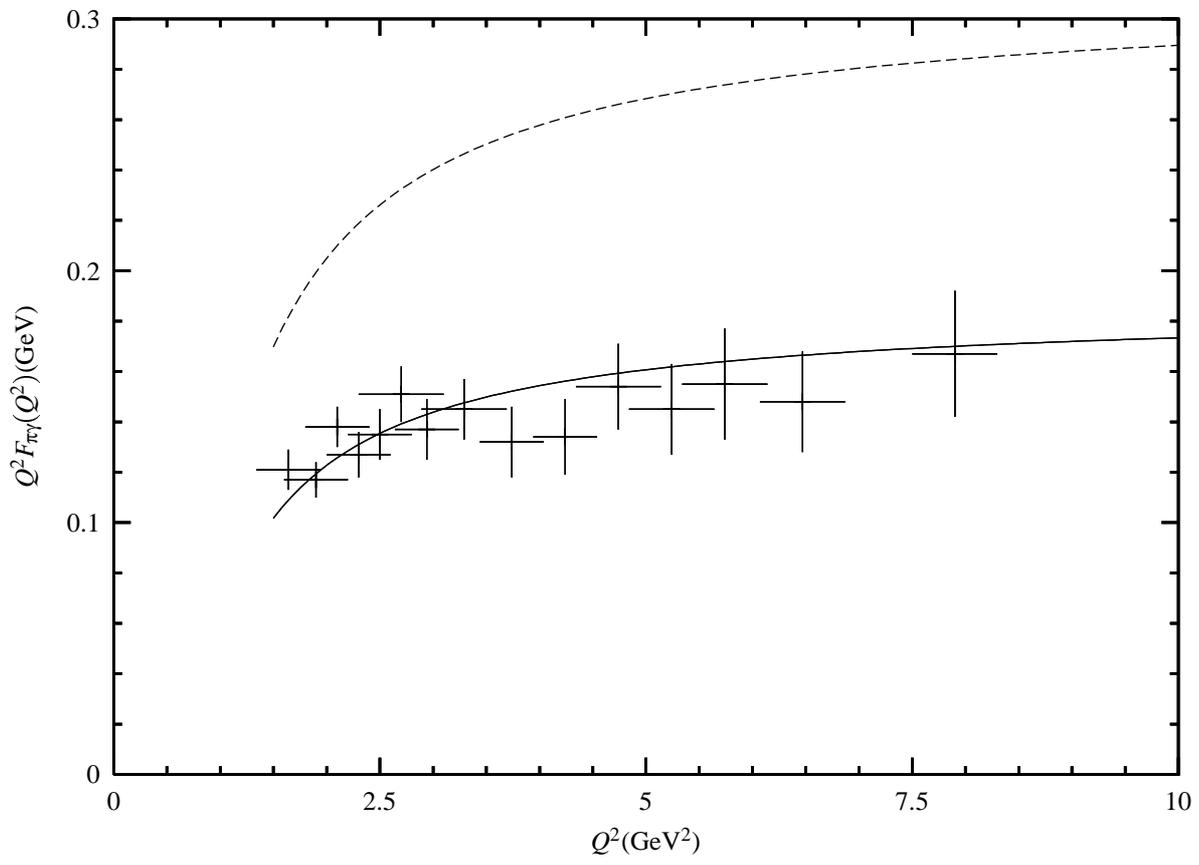
\end{document}